\documentstyle[editedvolume]{crckapb} 

\input epsf
\def\spose#1{\hbox to 0pt{#1\hss}}
\def\lta{\mathrel{\spose{\lower 3pt\hbox{$\mathchar"218$}}
     \raise 2.0pt\hbox{$\mathchar"13C$}}}
\def\gta{\mathrel{\spose{\lower 3pt\hbox{$\mathchar"218$}}
     \raise 2.0pt\hbox{$\mathchar"13E$}}}
\def\etal{{\it et al.\ }}
\begin{opening}
\title{Bright Star Clusters in the Antennae \protect\\
analysed with Evolutionary Synthesis}
\author{Uta Fritze - v. Alvensleben, Oliver Kurth}
\institute{Universit\"ats--Sternwarte, G\"ottingen, Germany}
\end{opening}
\runningtitle{Bright Star Clusters in the Antennae}
\begin{document}

\section{Introduction}
The Antennae galaxies (NGC 4038/39) are a pair of relatively nearby 
(${\rm D \sim 19~ Mpc}$ ${\rm for~H_0=75}$) gas-rich spirals of comparable 
mass in the process of merging with extensive dynamical and evolutionary 
synthesis modelling available. 
With HST WFPC1 
Whitmore \& Schweizer (1995) ({\bf WS95}) detected 738 bright Young Star 
Clusters ({\bf YSCs}) around the Antennae. 
YSCs have been detected in many interacting and starbursting galaxies, but the 
YSC system in the Antennae is the most populous one, well defining an 
exponential  Luminosity Function ({\bf LF}) over more than 5 mag 
above the completeness 
limit. Problems arising from crowding of the YSCs on a bright and variable 
galaxy background and from the PSF of WFPC1 may cause blending 
of YSCs and an overestimation of their effective radii. 

A question with far-reaching consequences, 
e.g. for the formation of elliptical galaxies, is if the YSCs formed in this 
interaction-triggered starburst are 
open or (proto-) globular clusters ({\bf GCs}). 

Concentration parameters involving the tidal radius  
being inaccessible to observations as a distinction criterion, interest 
focusses on the LF. While in the Milky Way and nearby galaxies 
the LF is exponential for open clusters but Gaussian 
for {\bf old} GC systems, {\bf the LF of a {\bf young} GC system is unknown}. 

So, the question, if GCs are formed in mergers gets related to the question 
as to the evolution of a GC system's LF. Vesperini (1997) shows that the strongly 
erodive dynamical effects on a GC system do {\bf not} change the shape 
of its mass function, provided it starts from a Gaussian. The aim of this paper 
is to study 
the effects of the spectrophotometric evolution of the YSC system on its LF. 

{\bf The formation of GCs is expected} in gas-rich spiral-spiral mergers 
on the basis of the high star formation efficiencies in
mergers and merger remnants 
(Fritze - v. Alvensleben \& Gerhard 1994a,b, Kurth 1996) together 
with hydrodynamical modelling that requires high star formation efficiences for 
GC formation (Brown \etal 1995) 

\section{Evolution of Star Clusters and Metallicities}
The evolution of star clusters for 5 different metallicities ${\rm 1 \cdot 
10^{-4} \leq Z \leq 0.04}$ is modelled as in Fritze - v. Alvensleben \& 
Burkert (1995, {\bf FB95}). Magnitudes, colours, stellar metallicity indicators, 
and synthetic spectra are calculated as a function of time and metallicity. 
Cluster metallicities can be predicted on the basis of the spiral progenitors' 
ISM abundances (Fritze - v. Alvensleben \& Gerhard 1994a). Our prediction of 
${\rm Z \sim 0.01}$ for NGC 7252, an Sc-Sc merger like the Antennae, is 
confirmed 
by spectroscopy of the brightest cluster W3 in NGC 7252 by Schweizer \& Seitzer 
(1993). For lack of spectroscopy and by analogy, we {\bf assume} ${\rm\ Z \sim 
0.01}$ also for the YSCs in the Antennae. 
Once cluster metallicities are known precise age dating from 
${\rm V-I ~ or ~U-V}$ colours becomes possible. 

{\bf Alone from their mean age of $\sim 1.3$ Gyr 
most of the YSCs in NGC 7252 are expected to be young GCs} (FB95). 

\section{Age Distribution of YSCs in the Antennae}
The mean ${\rm V-I}$ colour together with an average internal reddening 
correction results in a mean age of the YSC population in the Antennae of 
$0.2 \pm 0.2$ Gyr, in agreement with global starburst age and dynamical time 
since pericenter (Kurth 1996, Barnes 1988). 

Assuming a uniform age for the YSCs the time evolution of the LF simply is a shape-conserving 
shift towards fainter magnitudes. For a subsample of YSCs with effective radii 
${\rm R_{eff} \leq 10}$ pc Fritze - v. Alvensleben (1996) shows that their LF 
after $\sim 12$ Gyr of evolution becomes compatible with a ``normal'' GC LF at 
least up to the turnover, though 
with some overpopulation of the faintest bins. Since these are most susceptible 
to depopulation by dynamical destruction and since the colour distribution also 
agrees with its Milky Way counterpart she argues that -- at least for YSCs with 
${\rm R_{eff} \leq 10}$ pc the LF does not preclude them from being GCs 
(at variance with van den Bergh's (1995) statement for the entire YSC sample). 

In an ongoing starburst like in the Antennae, the age spread among star clusters 
may be comparable to their ages. Therefore we analyse the age distribution 
of the YSCs as derived from their individual observed colours in ${\rm V-I}$ 
and, as far as available, ${\rm U-V}$. Fig. 1 clearly reveals 
two distinct age populations: a large population of 399 (probably 481) YSCs with ages 
from 0 -- $2\cdot 10^8$ yr from the 
present burst and a small population of 32 (probably 69) {\bf old GCs} from the parent 
galaxies. The small number of interloopers is probably due to inhomogeneous 
internal reddening (see Fritze - v. 
Alvensleben 1998). Strikingly, the number of old GCs agrees with the number 
expected in case the Milky Way and M31 were merging instead of NGC 4038 and 4039. 

\begin{figure}
\epsfysize=6.0cm
\centerline{\epsffile{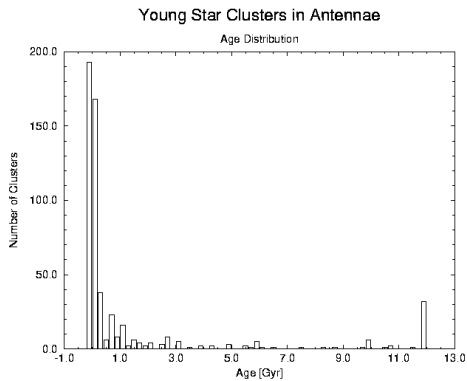}}
\vskip -1.0cm
\caption{Age distribution of bright star clusters in the Antennae.}
\vskip -0.35cm
\end{figure}

The average as well as the distribution of effective radii is not significantly different 
for the old GC and the YSC subsamples. Statistical tests assure the fraction of 
blended pairs to be
$< 10\%$ even among clusters with ${\rm R_{eff} > 10 ~pc}$ (see Fritze - v. Alvensleben 1998 for details). 

\section{Evolution of the LF of the YSCs in the Antennae}
Once we have individual ages of the YSCs our models also allow to calculate 
their individual fading until a common age of say 12 Gyr. Meurer (1995) 
already suspects that age spread effects might change the shape of the LF in 
the course of evolution. The brightest clusters tend to be the youngest and will 
therefore fade more, while part of the faintest clusters will fade less than 
average. We find that not only the LF of the subsample of YSCs with ${\rm R_{eff} 
\leq 10}$ pc evolves into a ``perfectly normal'' Gaussian, but also the LF 
of the {\bf entire YSC sample} (see Fig. 2). The turnover occurs at 
${\rm M_{V_o} \sim -6.9}$ mag, more than 1 mag brighter than the (evolved) 
completeness limit. The difference of $\sim 0.2$ mag to the typical elliptical 
galaxy GC systems'  ${\rm M_{V_o} \sim -7.1}$ mag (Harris 1991, Ashman \etal 1995) 
is a consequence of the enhanced metallicity of this secondary cluster 
population. 

\begin{figure}
\epsfysize=7.0cm
\centerline{\epsffile{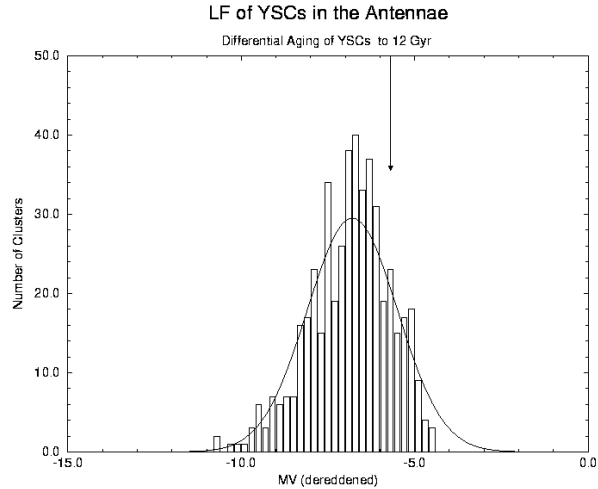}}
\vskip -0.5cm
\caption{LF of YSCs in the Antennae as calculated from individual ages together 
with the resulting individual fading until 12 Gyr for every cluster. 
A Gaussian with $\langle {\rm M_{V_0}} \rangle =-6.9$ mag and 
$\sigma({\rm M_{V_0)}})=1.3$ mag is overplotted, normalised to 
the number of clusters in the histogram. The vertical arrow marks the evolved 
observational completeness limit.}
\vskip -0.2cm
\end{figure}

{\bf We conclude:} Properly accounting for age spread effects and the resulting differences 
in fading the observed exponential LF of the entire YSC system around the merging Antennae 
galaxies evolves into a ``typical'' Gaussian GC LF over 12 Gyr 
(cf. Fritze -v. Alvensleben 1998). 

Together with Vesperini's (1997) result that dynamical effects 
destroying as much as 60 \% of an original GC population do not change its mass function 
if this initially had a Gaussian shape our results indicate that -- while there 
surely 
will be some open clusters among the YSC population -- the bulk of the YSCs in 
the Antennae may well be young GCs. Their number is large enough for the Antennae to 
transform into an elliptical with ``normal'' specific GC frequency. 

\smallskip\noindent
{\sl UFvA greatfully acknowledges a travel grant (Fr 916/4-1) from the DFG.}

\end{document}